\documentstyle[12pt]{article}

\csname @addtoreset\endcsname{equation}{section}

\def\bseq{\begin{subequation}}  
\def\eseq{\end{subequation}}
\def\bsea{\begin{subeqnarray}}  
\def\esea{\end{subeqnarray}}

\newcommand{\bbox}{\lower.2ex\hbox{$\Box$}}
\newcommand{\eqn}[1]{(\ref{#1})}

\newcommand{\ghe}[3]{$\stackrel{\textstyle #1}{\scriptstyle (#2,#3)}$}
\newcommand{\lre}{\multicolumn{3}{c}{$\stackrel{*}{\longleftrightarrow}$}}
\newcommand{\Ka}{K\"ahler}

\newcommand{\beq}{\begin{equation}}
\newcommand{\eeq}{\end{equation}}
\newcommand{\bea}{\begin{eqnarray}}
\newcommand{\eea}{\end{eqnarray}}
\newcommand{\ena}{\end{eqnarray}}

\newcommand {\non}{\nonumber}

\renewcommand{\a}{\alpha}
\renewcommand{\b}{\beta}

\renewcommand{\d}{\delta}

\newcommand{\pa}{\partial}
\newcommand{\g}{\gamma}

\newcommand{\e}{\epsilon}

\newcommand{\m}{\mu}

\newcommand{\n}{\nu}

\newcommand{\F}{\Phi}

\renewcommand{\P}{\Pi}

\newcommand{\s}{\sigma}
\renewcommand{\S}{\Sigma}
\renewcommand{\t}{\tau}

\newcommand{\Db}{\bar{D}}

\newcommand{\sigmab}{\bar{\sigma}}
\newcommand{\Sigmab}{\bar{\Sigma}}
\newcommand{\Phib}{\bar{\Phi}}

\newcommand{\ad}{{\dot{\alpha}}}
\newcommand{\bd}{{\dot{\beta}}}
\newcommand{\gd}{{\dot{\gamma}}}

\begin{document}

\begin{titlepage}
\begin{flushright} IFUM-579-FT\\ KUL-TF-97/26  \\ hep-th/9710166
\end{flushright}
\vfill
\begin{center}
{\LARGE\bf The nonminimal scalar multiplet: duality, sigma-model, 
beta-function}\\
\vskip 27.mm  \large
{\bf   Silvia Penati $^1$, Andrea Refolli $^1$,\\
Antoine Van Proeyen $^{2,\dagger}$ and  Daniela Zanon $^1$ } \\
\vfill
{\small
 $^1$ Dipartimento di Fisica dell'Universit\`a di Milano and\\
INFN, Sezione di Milano, via Celoria 16,
I-20133 Milano, Italy\\
\vspace{6pt}
$^2$ Instituut voor theoretische fysica,
Katholieke Universiteit Leuven,\\ B-3001 Leuven, Belgium. }
\end{center}
\vfill

\begin{center}
{\bf ABSTRACT}
\end{center}
\begin{quote}
We compute in superspace the one-loop beta-function for the nonlinear
sigma-model defined in terms of the nonminimal scalar multiplet.
The recently proposed quantization of this complex linear
superfield, viewed as the field strength of an unconstrained gauge spinor
superfield, allows to handle efficiently the infinite tower of ghosts
via the Batalin-Vilkovisky formalism. We find
that the classical duality of the nonminimal scalar and chiral multiplets
is maintained at the quantum one-loop level.
\vfill      \hrule width 5.cm
\vskip 2.mm
{\small
\noindent $^\dagger$ Onderzoeksdirecteur FWO, Belgium }
\end{quote}
\begin{flushleft}
October 1997
\end{flushleft}
\end{titlepage}

\section{Introduction}

It is well known that an alternative description of the scalar multiplet
is provided by the complex linear superfield \cite{b1}.
The equivalence of the two
formulations can be exhibited by means of a duality transformation which
relates the two multiplets to each other. Starting with the scalar
multiplet $\Phi$, with $\Db_\ad \Phi=0$, one can write
a first order action
\beq
S=- \int d^4x~ d^4 \theta ~ \left[ \Sigmab \Sigma + \S\Phi + \Sigmab
\Phib \right]
\eeq
where $\S$ is an auxiliary superfield.
Using the equations of motion
to eliminate the superfields $\S$, $\Sigmab$, one obtains
the usual chiral superfield
action. Eliminating instead the superfields  $\Phi$, $\Phib$,
whose equations of motion
impose the linearity constraint $\Db^2 \S = D^2 \Sigmab =0 $, leads
to the linear superfield action.
In the same manner one can start with the linear multiplet $\S$ satisfying
the constraint $\Db^2 \S =0$, and write the first order action
\beq
S= \int d^4x~ d^4 \theta ~ \left[ \Phib \Phi + \S\Phi + \Sigmab
\Phib \right]
\eeq
where now $\Phi$ has to be thought of as auxiliary and unconstrained.
We note that the duality transformation exchanges the field
equations with the constraints, as standard electric-magnetic
duality leads to the exchange of the Maxwell equations with
the Bianchi identities.

More generally, in terms of these multiplets, it is possible to
formulate supersymmetric $\s$-models.
We consider superfields
$\Phi^\m $, $\Phib^{\bar{\m}}$, $\S^\m $, $\Sigmab^{\bar{\m}}$ with
$\m,\bar{\m}=1,\dots,n$ and write the first order action
\beq
S= \int d^4x~ d^4\theta~ \left[ K(\Phib, \Phi) + \S \Phi +\Sigmab \Phib
\right]
\label{actionsigma}
\eeq
where $\S$ are linear superfields, $\Db^2 \S=0$, and $\Phi$ initially
unconstrained (we do not indicate the indices $\m$, $\bar{\m}$ on the
superfields in order to simplify the notation).
Varying with respect to $\S$ one obtains the chirality constraint on
$\Phi$, so that the quadratic terms, being total derivatives,
can be dropped and one is left with the standard $\s$-model action for
chiral superfields.
On the other hand the dual model is obtained once the variation with
respect to $\Phi$ is considered: the equations
\beq
\S=-\frac{\pa K}{\pa \Phi}\qquad \qquad \Sigmab=-\frac{\pa K}{\pa \Phib}
\eeq
have to be solved, $\Phi=\Phi(\S,\Sigmab)$, $\Phib=\Phib(\S,\Sigmab)$,
so that the dual action becomes
\beq
S =\int d^4x~ d^4\theta~ \tilde{K}(\S,\Sigmab)
\eeq
where $\tilde{K}$ is the Legendre transform of $K$
\beq
\tilde{K}(\S,\Sigmab)= [K(\Phi,\Phib)+\S\Phi+\Sigmab
\Phib]|_{\Phi=\Phi(\S,\Sigmab), \Phib=\Phib(\S,\Sigmab)}
\label{dualpot}
\eeq
Thus the duality transformations are implemented exchanging the potentials
\beq
K(\F,\Phib)\rightarrow \tilde{K}(\S,\Sigmab)
\eeq
At the level of the matrix given by the second derivatives
of the potential, i.e. defining
\beq
G=\left( \matrix{\frac{\pa^2 K}{\raise-.04cm\hbox{$\scriptstyle
\pa \Phi \pa \Phib$}}&
\frac{\pa^2 K}{\raise-.04cm\hbox{$\scriptstyle
\pa \Phi \pa \Phi$} } \cr
~~&~~ \cr
\frac{\pa^2 K}{\raise-.04cm\hbox{$\scriptstyle
\pa \Phib \pa \Phib$}} &
\frac{\pa^2 K}{\raise-.04cm\hbox{$\scriptstyle
\pa \Phib \pa \Phi$}} \cr}
\right)
\eeq
and similarly for $\tilde{G}(\S,\Sigmab)$,
it is easy to show that
\beq
G(\Phi,\Phib)\rightarrow \tilde{G}(\S,\Sigmab) = - G(\Phi,\Phib)^{-1}
\label{dualmetric}
\eeq
(These transformations generalize the ones valid for bosonic
$\s$-models with an isometry \cite{bu}.)

The duality properties of these theories are well understood in
a superfield formulation. On the other hand the
situation is not so clear in terms of component fields.
The two multiplets have
the same content of physical degrees of freedom, but they differ
in their auxiliary field structure and it appears that
the elimination of these auxiliary fields via their equations
of motion might lead to $\s$-models quite different in the two cases \cite{b2}.
Therefore in attempting to understand duality issues at the
quantum level it seems safer to stick to a superspace
approach. While for the chiral superfield the quantization
is not an issue, a direct quantum formulation of the complex
linear superfield is not known. Indeed we are able to
perform functional integration and differentiation in
chiral superspace \cite{b1}, but we do not have a corresponding setup
in the linear case. One way out is to solve the linearity
constraint $\Db^2 \S =0$ in terms of unconstrained gauge
superfields $\s^\a$, $\bar{\s}^\ad$ whose quantization however
leads to an infinite tower of ghosts. In a recent paper
\cite{b3} a complete solution of the problem
has been obtained by using the Batalin-Vilkovisky \cite{b4} approach
to gauge fix the infinite sequence of
invariances. Although conceptually straightforward
the method looks complicated and any practical application
difficult to envisage. One aim of the
present paper is to prove that some of these difficulties
are not real obstacles: we adopt the recently proposed
 quantization procedure
and investigate the quantum duality properties of the complex linear
$\s$-model. Since these theories are not
renormalizable in four dimensions, in the following we restrict our
attention to the two-dimensional situation.

At the classical level the chiral and the linear
$\s$-models defined on dual backgrounds represent different
parametrizations of the same theory.
The manipulations that bring one theory into
the other are essentially based on functional integrations
performed in a different order. Consequently in order to
address quantum issues, the obvious question to ask is
about quantum duality of the properly regularized theories.
The simplest object which directly relates to the renormalization
properties of a given model, is the $\b$-function.
We compute the one-loop $\b$-function for the nonlinear $\s$-model
in terms of the complex linear superfields and compare it with the
well known corresponding result \cite{b5,b6} for the $\s$-model in terms
of chiral superfields.

In the next section we give a concise description of the
quantum-background field approach, best suited for perturbative
calculations. In section 3 we show how to treat the infinite tower of
fields that enter in the quantization of the linear superfield and
how to obtain an effective propagator to be used in section 4, where
the one-loop $\b$-function is explicitly computed.
We conclude with some comments and problems open to future
investigations. Few useful identities are collected in an appendix.

\section{The background field method}

First we briefly review the situation for the nonlinear
$\s$-model defined in terms of chiral superfields \cite{b6}. 
The action can be written as
\beq
S=\int d^2x~d^4\theta~K(\Phi, \Phib)
\label{actionchiral}
\eeq
where $\Phi^\m$, $\Phib^{\bar{\m}}$ are interpreted as complex
coordinates of a \Ka\ manifold.
In order to perform perturbative calculations one shifts the
superfields with a linear quantum-background splitting
\beq
\Phi\rightarrow \Phi+\Phi_0 \qquad, \qquad
\Phib\rightarrow \Phib+\Phib_0
\eeq
and expand the action around the background $\Phi_0$, $\Phib_0$.
One separates the free kinetic action of the quantum fields
from the interaction vertices
\bea
S&=&\int d^2x~d^4\theta~ \left[\Phi^\m \Phib^{\bar{\n}}\d_{\m \bar{\n}}+
[K_{\m \bar{\n}}(\Phi_0,\Phib_0)
-\d_{\m \bar{\n}}]\Phi^\m \Phib^{\bar{\n}} \right. \non\\
&~&+\left.\frac{1}{2}K_{\m \n}(\Phi_0,\Phib_0)\Phi^\m \Phi^{\n}
+\frac{1}{2}K_{\bar{\m} \bar{\n}}(\Phi_0,\Phib_0)\Phib^{\bar{\m}}
\Phib^{\bar{\n}}+
\dots \right]
\label{actionquad}
\eea
where
\beq
K_{\m,\dots,\bar{\m},\dots}\equiv \frac{\pa}{\pa\Phi^\m}\dots
\frac{\pa}{\pa \Phib^{\bar{\m}}}\dots K(\Phi,\Phib)
\eeq
Quantum calculations can be performed using superspace
Feynman diagrams and conventional $D$-algebra techniques.
The quadratic action in (\ref{actionquad}) is sufficient for
one-loop computations. Using dimensional regularization
in $2-2\e$ dimensions, the one-loop divergent contribution
to the \Ka\ potential is given by
\beq
K^{(1)}= \frac{1}{\e} \rm{tr} \log K_{\m \bar{\n}}
\label{1loopdiv}
\eeq
The corresponding renormalization of the \Ka\ metric is obtained
in terms of the one-loop $\b$-function
\beq
\b^{(1)}_{\m \bar{\m}}=-\pa_\m \pa_{\bar{\m}} \rm{tr} \log K_{\n \bar{\n}}
=-R_{\m \bar{\m}}
\eeq
being $R_{\m \bar{\m}}$ the Ricci tensor of the manifold.

In order to study the corresponding problem for the
$\s$-model written in terms of complex linear superfields,
we start with the general superspace action
\beq
S=\int d^2x~ d^4\theta ~ F(\Sigma, \Sigmab)
\label{actionlinear}
\eeq
where $\Db^2 \S = D^2 \Sigmab =0$. As we have mentioned above, if
$F=\tilde{K}$ with $\tilde{K}$ given in (\ref{dualpot})
the two models described by the actions in (\ref{actionlinear}) and
(\ref{actionchiral}) are classically dual to each other.
Now we want to step up to the quantum level.

We follow the same approach
described for the nonlinear
$\s$-model in terms of chiral superfields
 and use the background field method in perturbation
theory, shifting
\beq
\S\rightarrow \S+\S_0 \qquad, \qquad
\Sigmab\rightarrow \Sigmab+\Sigmab_0
\eeq
in (\ref{actionlinear}). As in the previous case the quantum fields
appear explicitly, while the background
dependence is contained in the interaction vertices through
derivatives  of the $F$ potential
\bea
S&=&\int d^2x~d^4\theta~ \left[-\S^\m \Sigmab^{\bar{\n}}\d_{\m \bar{\n}}+
[F_{\m \bar{\n}}(\S_0,\Sigmab_0)
+\d_{\m \bar{\n}}]\S^\m \Sigmab^{\bar{\n}} \right. \non\\
&~&+\left.\frac{1}{2}F_{\m \n}(\S_0,\Sigmab_0)\S^\m \S^{\n}
+\frac{1}{2}F_{\bar{\m} \bar{\n}}(\S_0,\Sigmab_0)\Sigmab^{\bar{\m}}
\Sigmab^{\bar{\n}}+
\dots \right]
\label{actionquad2}
\eea
with the definitions
\beq
F_{\m,\dots,\bar{\m},\dots}\equiv \frac{\pa}{\pa\S^\m}\dots
\frac{\pa}{\pa \Sigmab^{\bar{\m}}}\dots F(\S,\Sigmab)
\eeq
Up to this point we have repeated the same steps as in the
chiral superfield example. In the present case however, we have to deal
with the fact that a direct superspace quantization of the complex linear
superfield is not available. As explained in ref. \cite{b3}, one
way to proceed is to solve the linearity constraint
$\S=\Db_\ad \bar{\s}^\ad$, $\Sigmab=D_\a \s^\a$ in terms of
unconstrained spinor superfields $\s^\a$, $\bar{\s}^\ad$.
The gauge invariance introduced in this manner needs to be fixed
and the quantization gives rise to an infinite tower of
ghosts.
The Batalin-Vilkovisky method provides a systematic
procedure for obtaining the gauge-fixed action.
Due to the appearance of an infinite number of ghost fields
the final result looks rather intricate and difficult to use
in applications. In fact we show here that
choosing gauge-fixing functions flat (i.e. independent) with
respect to the background external fields, all the ghosts
essentially decouple. More precisely, the canonical
transformations, which in the Batalin-Vilkovisky approach are
necessary to go from the classical basis to the gauge-fixed basis,
produce nondiagonal terms between the quantum gauge
spinors $\s^\a$, $\bar{\s}^\ad$ and some of the ghost fields.
In the following we perform explicitly
the diagonalization which leads to the relevant
kinetic terms needed for the evaluation of the one-loop
$\b$-function.
As a non trivial check of the consistency of our calculation
we compute in a general gauge and prove that the resulting
physics (e.g. the $\b$-function) is independent of the
gauge parameters.

\section{Diagonalization and effective propagator}

We express the quantum linear superfields, viewed as the field
strength of some gauge spinors, in terms of $\s^\a$, $\bar{\s}^\ad$
so that the classical kinetic term in (\ref{actionquad2}) becomes
\beq
S_{cl}= -\int d^2x~d^4\theta~\bar
\sigma^{\dot \alpha}\bar D_{\dot \alpha} D_\alpha\sigma^\alpha\ 
\label{Scl0}
\end{equation}
In ref. \cite{b3} the gauge invariances, i.e. the infinite chain of
transformations with zero modes which  are responsible for the appearance
of a tower of ghosts, have been discussed in detail. We refer the
reader to that paper for the complete derivation of the Batalin-Vilkovisky
quantization of the model. Here we review only the few steps which are
relevant for our successive calculations.

The quantization of the classical action in (\ref{Scl0}) starts by
defining the minimal extended action
\beq
S_{min}=S_{cl}+\Phi_A^* \d \Phi^A
\label{Smin}
\eeq
where generically $\Phi_A$, $\d\Phi_A$ denote a field and its
gauge variation, while $\Phi^*_A$ denotes the corresponding antifield.
Then one adds to (\ref{Smin}) non-minimal fields and constructs
the extended action in classical basis. Gauge fixing is
performed by canonical transformations through the introduction of
a gauge fermion $\Psi(\Phi)$
\beq
\Phi_A^* \rightarrow \Phi_A^*+\frac{\d\Psi}{\d\Phi^A}
\label{cantransf}
\eeq
It is the iterative succession of canonical transformations
which is responsible for the mixing of the various fields and
ultimately requires a nontrivial diagonalization. Now we
outline the procedure on our specific model.

As mentioned above, since the gauge fixing does not introduce
any explicit coupling with the external background fields, we need only
consider those contributions in the nonminimal action and in the gauge
fermion which give rise to mixed, nondiagonal terms between the
ghosts and the quantum fields $\s^\a$,
$\bar{\s}^\ad$. This amounts saying that in the pyramid of
fields produced by the Batalin-Vilkovisky construction,
given schematically in table~\ref{tbl:fields4} for the
first four levels
($A_i$ is an abbreviation for the symmetrized set of indices
$(\a_1 \dots \a_i))$, it suffices to concentrate on the
fields and antifields of ghost number zero, which are those
in the upper left diagonal. Therefore in the
following we systematically ignore the remaining fields:
while important for the complete Batalin-Vilkovisky
construction, they do not play any role for the determination
of the effective propagator which will enter in the
$\b$-function calculation.
\tabcolsep 1pt
\begin{table}[htb]\caption{Fields up to fourth level. The first column
gives the level, and the second one indicates whether the fields are
bosonic or fermionic. The ghost numbers
of the field and of the antifield are also indicated.}
\label{tbl:fields4}\begin{center}\begin{tabular}{cccccccccccccccccc}
0~~~~&F &   &   &   &   &   &   & &\ghe{\sigma^{\alpha}}0{-1}&   &   &   &   &   
&   &
 &  \\
& &  &   &   &    &   &   &$\swarrow$ & &    &   &   &   &   &   &   &  \\
1~~~& F &  &      &   &   &   & \ghe{b_{\alpha}^{\dot \alpha}}{-1}0  &   &   &
&\ghe{\sigma^{\alpha_1\alpha_2}}1{-2}&   &
 &   &   &&   \\
& &  &      &   &   & $\swarrow$   &   &   &   & $\swarrow$   &   &   &   &
 &   &  &    \\
2~~~~& F &  &   &   &    \ghe{d^{A_2}_{\dot \alpha}}0{-1}  &
\lre     & \ghe{b_{A_2}^{\dot \alpha}}{-2}1  &
 &   &   &\ghe{\sigma^{A_2\alpha_3}}2{-3}     &   &   &\\
&B &  &   &   &    $\nu$  &
\lre     &$ \mu $ &
 &   &   &$\lambda $ &   &   &\\
& &  &   &   $\swarrow$&   &   &   & $\swarrow$   &   &   &   & $\swarrow$
 &   &      & &   & \\
3~~~~& F &  &
\ghe{e_{A_2}^{\dot A_2}}{-1}0 &
&   &   & \ghe{d^{A_3}_{\dot \alpha},d}1{-2}
&  \lre
 & \ghe{b_{A_3}^{\dot \alpha},b}{-3}2  &
 &   &   & \ghe{\sigma^{A_3\alpha_4},\varsigma }3{-4}  &  & \\
&B &  &
$ \rho^{\dot \alpha}$ &
&   &   & $\nu^{\alpha}$
&  \lre
 & $\mu_{\alpha}$ &
 &   &   & $\lambda^{\alpha}$  &  & \\
 &  &$\swarrow$&   &   &   & $\swarrow$   &   &   &   &$\swarrow$    &
 & &   & $\swarrow$ &  &   &     \\
4~~~~&\ghe{f^{A_3}_{\dot A_2} }0{-1}& \lre   &
\ghe{e_{A_3}^{\dot A_2} }{-2}1 &
&   &   & \ghe{d^{A_4}_{\dot \alpha},d^{\alpha}}2{-3}
&  \lre
 & \ghe{b_{A_4}^{\dot \alpha},b_{\alpha}}{-4}3  &
 &   & \ghe{\sigma^{A_4\alpha_5}
 ,\varsigma^{\alpha} }4{-5}      \\
&$\tau^{\alpha}_{\dot \alpha}$& \lre   &
$\rho^{\dot \alpha}_{\alpha}$&
&   &   & $\nu^{\alpha_1\alpha_2}$
&  \lre
 &$ \mu_{\alpha_1\alpha_2}$  &
 &   &  $\lambda^{\alpha_1\alpha_2}$     \\
\end{tabular}\end{center}\end{table}     \tabcolsep 6pt

In order to complete
\eqn{Scl0} to an invertible kinetic term one introduces gauge-fixing functions
\beq
F_\a^\ad = D_\a \sigmab^\ad ~~~~,~~~~~\bar{F}_\ad^\a = \Db_\ad \s^\a
\eeq
and corresponding (antighost) fields
$b_\alpha^{\dot \alpha}$ and their complex conjugates $\bar b_{\dot
\alpha}^\alpha$.  Indicating
the antifields of the antighosts respectively by
$b^{*\alpha}_{\dot \alpha}$ and $\bar b^{*\dot \alpha}_\alpha$, we
add  to $S_{min}$ the non--minimal term
(we do not indicate the integration
symbol and an overall minus sign, $-\int d^2x d^4\theta$, in the
definition of the action )
\begin{equation}
S_{nm,1}= \bar b^{*\dot
\alpha}_\alpha b^{*\alpha}_{\dot \alpha}\ 
\label{Snm1}
\end{equation}
Then we perform a canonical transformation generated by the gauge fermion
\begin{equation}
\Psi_1= k[b_\alpha^{\dot \alpha} \bar D_{\dot
\alpha}\sigma^\alpha +\bar
\sigma^{\dot \alpha}D_\alpha\bar b_{\dot \alpha}^\alpha ]\ 
\label{Psi1}
\end{equation}
$k$ being the gauge parameter\footnote{For simplicity we choose
 $k$ and all further
gauge parameters real.}
(note that in ref. \cite{b3} $k$ was chosen equal to one).
This implies the substitutions
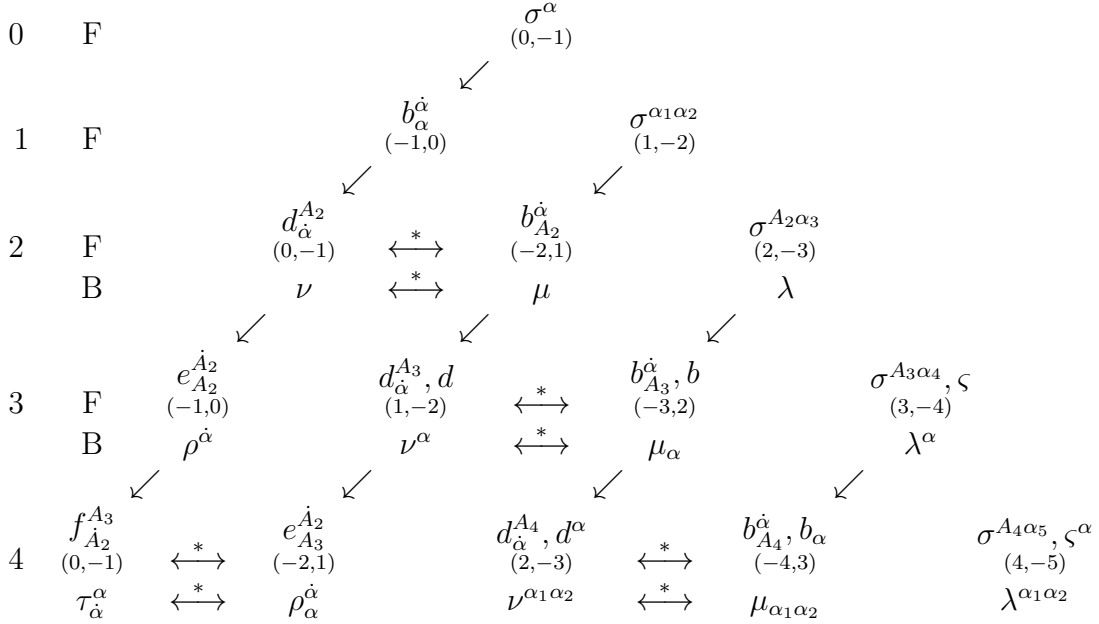
\bea
b^{*\alpha}_{\dot \alpha} &\rightarrow&
b^{*\alpha}_{\dot \alpha} +k\bar D_{\dot \alpha}
\sigma^\alpha\nonumber\\
\bar b^{*\dot \alpha}_\alpha &\rightarrow&
\bar b^{*\dot \alpha}_\alpha + k\bar \sigma^{\dot \alpha}D_\alpha
\label{can1}
\eea
We then obtain from the non-minimal action
\beq
S_1\rightarrow k\bar b^{*\dot
\alpha}_\alpha \bar D_{\dot \alpha} \sigma^\alpha + k \bar
\sigma^{\dot \alpha} D_\alpha b^{*\alpha}_{\dot \alpha} +  k^2\bar
\sigma^{\dot \alpha} D_\alpha \bar D_{\dot \alpha} \sigma^\alpha \
\label{nmact1}
\eeq
The last term combines with $S_{cl}$
and leads to a quadratic gauge-fixed action 
\beq
S_{Q,1}= \bar \sigma^{\dot
\alpha}[\Db_\ad D_\a +k^2 D_\a \Db_\ad]\sigma^\alpha
\label{SQ1}
\eeq
The rest gives
\beq
S_{*,1}=  \bar b^{*\dot \alpha}_\alpha b^{*\alpha}_{\dot \alpha}+
k\bar b^{*\dot \alpha}_\alpha \bar D_{\dot \alpha}  \sigma^\alpha +
k \bar \sigma^{\dot \alpha} D_\alpha b^{*\alpha}_{\dot \alpha} \
\label{Slevel1}
\eeq
Continuing, after the first level gauge-fermion in
\eqn{Psi1} we introduce
\begin{equation}
\Psi_2=b_\alpha^{\dot \alpha}
\left( D_\beta d^{(\beta\alpha)}_{\dot \alpha}+
i\partial^\alpha{}_{\dot
\alpha}\nu \right) +\left(\bar{d}_\a^{(\ad\bd)} \Db_\bd
-\bar{\n} i \pa_\a^{~\ad}\right) \bar{b}^\a_\ad
+ \dots
\label{Psi2}
\end{equation}
where we have considered only
the part which is relevant for our present discussion.
Note that the gauge fermion at odd levels connects fields of that
level with fields of the previous level, already present in the
action constructed that far: in this case the gauge parameter cannot be
rescaled. At even levels instead one connects fields of that level with
fields below, not yet present in the action:
therefore appropriate field redefinitions of the new fields allow
to rescale some of the gauge parameters to one. In (\ref{Psi2})
and in the following, at every even level,
we implicitly assume to have performed such rescalings.

The canonical transformations induced by (\ref{Psi2}) are given by
\bea
b^{*\a}_\ad &\rightarrow& b^{*\a}_\ad +D_\b d_\ad^{(\b\a)} +
i \pa ^\a_{~\ad} \n \nonumber\\
\bar b^{*\dot \alpha}_\alpha &\rightarrow&
\bar b^{*\dot \alpha}_\alpha +
\bar{d}_\a^{(\ad\bd)} \Db_\bd -\bar{\n} i \pa_\a^{~\ad}
\label{can2}
\eea
The action receives accordingly the following contributions
\beq
S_2 \rightarrow k \bar{\s}^\ad D_\a i \pa ^\a_{~\ad} \n -
k \bar{\n} i \pa_\a^{~\ad} \Db_\ad \s^\a
+\left[ \bar{d}_\a^{(\ad\bd)} \Db_\bd -\bar{\n} i \pa_\a^{~\ad}\right]
\left[ D_\b d_\ad^{(\b\a)} + i \pa ^\a_{~\ad} \n\right]
\label{nmact2}
\eeq
Clearly, as we had anticipated, we have produced mixed (non diagonal)
terms between
the various fields, $\s^\a$, $\n$ and $d_\ad^{(\a\b)}$.
Note however that mixed terms between the $\s^\a$'s and the ghosts
$d_\ad^{(\a\b)}$ do not arise for symmetry reasons (e.g.
$\bar{\s}^\ad D_\a D_\b d_\ad^{(\b\a)}=0$).
This pattern repeats itself at every step.

We proceed introducing the gauge fermion
\beq
\Psi_3= k_1 e^{(\ad\bd)}_{(\a\b)}
\Db_\bd d^{(\a\b)}_\ad +k_1\bar{d}^{(\ad\bd)}_\a
D_\b \bar{e}^{(\a\b)}_{(\ad\bd)}+
k'_1\rho^{\dot \alpha}\bar D_{\dot \alpha}\nu
-k'_1 \bar{\n} D_\a \bar{\rho}^\a + \dots
\label{Psi3}
\eeq
with canonical transformations
\bea
\rho^*_\ad &\rightarrow& \rho^*_\ad +k'_1 \Db_\ad \n \nonumber\\
\bar{\rho}^*_\a &\rightarrow& \bar{\rho}^*_\a- k'_1\bar{\n} D_\a \nonumber\\
{e}^{*(\a\b)}_{(\ad\bd)}&\rightarrow& {e}^{*(\a\b)}_{(\ad\bd)}
+k_1\Db_{(\bd} d^{(\a\b)}_{\ad)} \nonumber\\
\bar{e}^{*(\ad\bd)}_{(\a\b)}&\rightarrow & \bar{e}^{*(\ad\bd)}_{(\a\b)}
+k_1\bar{d}^{(\ad\bd)}_{(\a} D_{\b)}
\label{can3}
\eea
These shifts affect the nonminimal part of the action at the third
level
\beq
S_{nm,3}= \bar{e}^{*(\ad\bd)}_{\a\b} e^{*(\a\b)}_{\ad\bd}
-\bar{\rho}^*_\a i\pa^{\a\ad} \rho^*_\ad
\label{Snm3}
\eeq
in the following way
\bea
S_3 &=& k^{'2}_1 \bar{\n} D_\a i \pa^{\a\ad} \Db_\ad \n
-\bar{\rho}^*_\a i\pa^{\a\ad} \rho^*_\ad
 -k'_1 \bar{\rho}^*_\a i\pa^{\a\ad}\Db_\ad \n
+k'_1 \bar{\n} D_\a i \pa^{\a\ad} \rho^*_\ad \nonumber\\
&~&+k_1^2 \bar{d}^{(\ad\bd)}_{(\a} D_{\b)}
\Db_{(\bd} d^{(\a\b)}_{\ad)} +\bar{e}^{*(\ad\bd)}_{(\a\b)}
e^{*(\a\b)}_{(\ad\bd)}+k_1\bar{d}^{(\ad\bd)}_{(\a} D_{\b)}
e^{*(\a\b)}_{(\ad\bd)}+k_1\bar{e}^{*(\ad\bd)}_{(\a\b)}
\Db_{(\bd} d^{(\a\b)}_{\ad)}\nonumber\\
&~&~~~~~~~~~~
\label{nmact3}
\eea
At the fourth level, in complete analogy with the choice made at the second
level in (\ref{Psi2}), we introduce the gauge fermion
\bea
\Psi_4&=&e^{(\ad\bd)}_{(\a\b)} \left( D_\g f^{(\a\b\g)}_{(\ad\bd)}
+i\pa^\a_{~\ad} \t^\b_{~\bd}\right)
+\left(\bar{f}^{(\ad\bd\gd)}_{(\a\b)} \Db_\gd
-\bar{\t}^\bd_{~\b} i\pa^{~\ad}_\a\right) \bar{e}^{(\a\b)}_{(\ad\bd)}
\nonumber\\
&&+ k^{''}_1\rho^\ad D_\a \t^\a_{~\ad}
-k^{''}_1\bar{\t}_\a^{~\ad} \Db_\ad \bar{\rho}^\a
+\dots
\label{Psi4}
\eea
(Note that we have rescaled the fields $f^{(\a\b\g)}_{(\ad\bd)}$ and
$\t^\b_{~\bd}$ to set equal to one two of the three gauge parameters
that should appear in (\ref{Psi4}).)
Now we have to perform the substitutions
\bea
\rho^*_\ad &\rightarrow & \rho^*_\ad + k^{''}_1D_\a \t^\a_{~\ad} \nonumber\\
\bar{\rho}^*_\a &\rightarrow & \bar{\rho}^*_\a - k^{''}_1\bar{\t}_
\a^{~\ad} \Db_\ad \nonumber\\
e^{*(\a\b)}_{(\ad\bd)}&\rightarrow & e^{*(\a\b)}_{(\ad\bd)} +
D_\g f^{(\a\b\g)}_{(\ad\bd)}+i\pa^{(\a}_{~(\ad} \t^{\b)}_{~\bd)} \nonumber\\
\bar{e}^{*(\ad\bd)}_{(\a\b)}&\rightarrow & \bar{e}^{*(\ad\bd)}_{(\a\b)}+
\bar{f}^{(\ad\bd\gd)}_{(\a\b)} \Db_\gd-\bar{\t}^{(\bd}_{~(\b} 
i\pa^{~\ad)}_{\a)}
\label{can4}
\eea
When inserted in (\ref{nmact3}) the shifts in
(\ref{can4}) give rise to
\bea
S_4& \rightarrow& k'_1 k^{''}_1\bar{\t}_\a^{~\ad} i\pa^\a_{~\ad} \Db^2 \n
-k'_1 k^{''}_1 \bar{\n}i\pa_\a^{~\ad} D^2 \t^\a_{~\ad}
+k^{''2}_1\bar{\t}_\a^{~\bd} \Db_\bd i\pa^{\a\ad} D_\b \t^\b_{~\ad}
\nonumber\\
&&+\left(\bar{f}^{(\ad\bd\gd)}_{(\a\b)} \Db_\gd
-\bar{\t}^\bd_{~(\b} i\pa^{~\ad}_{\a)} \right)
\left( D_\g f^{(\a\b\g)}_{(\ad\bd)}+i\pa^\a_{~(\ad} \t^\b_{~\bd)}\right)
\label{nmact4}
\eea
Here again we see new non diagonal terms between the ghosts
$\n$ and $\t^\a_{~\ad}$, but no crossed terms
between the $d^{(\a\b)}_\ad$'s and the $f^{(\a\b\g)}_{(\ad\bd)}$'s.
This is a general feature which can be easily implemented at
higher levels given the structure of the symmetrized
indices on the various fields. In fact at this point
it is simple to understand how the story continues and one can easily write
all at once the infinite sum of terms that enter the quadratic part of the
gauge-fixed action. To this end it is convenient to
introduce a compact notation and rename appropriately the
 fields in the upper left diagonal of table~\ref{tbl:fields4}.
 At even levels we call the fermionic fields
$d^{A_{n+1}}_{\dot{A}_n}$
and the bosonic fields
 $\n^{A_{n-1}}_{\dot{A}_{n-1}}$, where $n=0,1,\dots$ (but $\nu$ does not
 exist for $n=0$),
 so that, for example, at zero level
we have set $d^\a\equiv \s^\a$, $\bar{d}^\ad\equiv
\bar{\s}^\ad$ and so on. At odd levels in the same way, with
obvious identifications with respect to the fields in table 1, we
call the fermionic fields $e^{\dot{A}_n}_{A_n}$, the bosonic fields
$\rho^{\dot{A}_{n-1}}_{A_{n-2}}$, $n=1,2,\dots$ (the field $\rho$
does not exist for $n=1$). Thus we can write
the classical action and the non-minimal terms relevant for our
purpose as
\beq
S_{cl}+S_{nm}=\bar{\s}^\ad \Db_\ad D_\a \s^\a
+\sum_{n=0}^{\infty} \bar{e}^{*{\dot{A}_{n+1}}}_{A_{n+1}}
e^{*A_{n+1}}_{\dot{A}_{n+1}}
-\sum_{n=1}^{\infty} \bar{\rho}^{*{\dot{A}_{n-1}}}_{\a A_{n-1}}
 i\pa^{\a\ad} \rho^{*{A_{n-1}}}_{\ad \dot{A}_{n-1}}
\label{Sgen}
\eeq
Using the same notation we obtain the total gauge fermion
\bea
\Psi&=&\sum_{n=0}^{\infty}k_n e^{(\ad\dot{A}_n)}_{A_{n+1}}
\Db_\ad d^{A_{n+1}}_{\dot{A}_n}+\sum_{n=1}^{\infty}k'_n
\rho^{(\ad\dot{A}_{n-1})}_{A_{n-1}}\Db_\ad
\n^{A_{n-1}}_{\dot{A}_{n-1}} \nonumber\\
&&+\sum_{n=0}^{\infty} e^{\dot{A}_{n+1}}_{A_{n+1}}
D_\b d^{(\b A_{n+1})}_{\dot{A}_{n+1}}
+\sum_{n=0}^{\infty} e^{(\ad\dot{A}_n)}_{(\a A_n)}
i\pa^\a_{~\ad} \n^{A_n}_{\dot{A}_n}
+\sum_{n=1}^{\infty} k^{''}_n \rho^{\dot{A}_n}_{A_{n-1}}D_\a
\n^{(\a A_{n-1})}_{\dot{A}_n}\nonumber\\
&&+\sum_{n=0}^{\infty}k_n \bar{d}^{\dot{A}_{n+1}}_{A_n}D_\a
\bar{e}^{(\a A_n)}_{\dot{A}_{n+1}} -\sum_{n=1}^{\infty}k'_n
\bar{\n}^{\dot{A}_{n-1}}_{A_{n-1}} D_\a
\bar{\rho}^{(\a A_{n-1})}_{\dot{A}_{n-1}}\nonumber\\
&&+\sum_{n=0}^{\infty} \bar{d}^{(\bd \dot{A}_{n+1})}_{A_{n+1}}
\Db_\bd \bar{e}^{A_{n+1}}_{\dot{A}_{n+1}}
-\sum_{n=0}^{\infty}\bar{\n}^{\dot{A}_n}_{A_n} i\pa_\a^{~\ad}
\bar{e}^{(\a A_n)}_{(\ad \dot{A}_n)}
-\sum_{n=1}^{\infty} k^{''}_n \bar{\n}^{(\ad\dot{A}_{n-1})}_{A_n}
\Db_\ad \bar{\rho}^{A_n}_{\dot{A}_{n-1}}
\label{fermgen}
\eea
One can easily check that the terms in (\ref{Psi1}) and (\ref{Psi2})
correspond to the $n=0$ contributions in (\ref{fermgen}), while the
gauge fermions at the third and fourth level in (\ref{Psi3}) and
(\ref{Psi4}) are reproduced by the $n=1$ coefficients of the various sums
in the above expression. The gauge fermion in (\ref{fermgen})
implies canonical transformations
that, once inserted in (\ref{Sgen}), reconstruct the quadratic
gauge-fixed action
\bea
S_{gf}&=& \sum_{n=0}^{\infty}\left[ \bar{d}_{A_n}^{(\bd \dot{A}_n)}
\left(\Db_\bd D_\b + k_n^2 D_\b \Db_\bd \right)
d^{(\b A_n)}_{\dot{A}_n}+k_n \bar{d}^{(\bd \dot{A}_n)}_{(A_n}
D_{\a_{n+1})} i\pa^{\a_{n+1}}_{~\bd} \n^{A_n}_{\dot{A}_n}\right.\nonumber\\
&&+\bar{d}^{(\bd\gd\dot{A}_n)}_{(\g A_n)} \Db_\bd i\pa^{\g}_{~\gd}
\n^{A_n}_{\dot{A}_n}
-k_n\bar{\n}^{\dot{A}_n}_{A_n} i\pa_{\b}^{~\ad_{n+1}}
\Db_{(\ad_{n+1}} d^{(A_n \b)}_{\dot{A}_n)}\nonumber\\
&& \left. -\bar{\n}^{\dot{A}_n}_{A_n} i\pa_{\g}^{~\gd} D_\b
d^{(\b\g A_n)}_{(\gd \dot{A}_n)}
-\bar{\n}^{(\dot{A}_n}_{(A_n} \left( i\pa_{\b)}^{~\bd)}
i\pa^{\b}_{~\bd}-k^{'2}_{n+1} D_{\b)} i \pa^{\bd) \b} \Db_\bd
\right) \n^{A_n}_{\dot{A}_n}\right] \nonumber\\
&&+\sum_{n=1}^{\infty} \left[ \frac{1}{n}k'_n k^{''}_n
\bar{\n}^{(\bd \dot{A}_{n-1})}_{(\a A_{n-1})} \Db^2 i\pa^\a_\bd
\n^{A_{n-1}}_{\dot{A}_{n-1}} -
\frac{1}{n} k'_n k^{''}_n \bar{\n}^{\dot{A}_{n-1}}_{A_{n-1}}
D^2 i\pa_\b^{~\ad} \n^{(\b A_{n-1})}_{(\ad \dot{A}_{n-1})}\right. \nonumber\\
&&~~~~~~~~\left. +k^{''2}_n \bar{\n}^{(\bd \dot{A}_{n-1})}_{(\a A_{n-1})}\Db_\bd
i\pa^{\a\ad} D_\b \n^{(\b A_{n-1})}_{(\ad \dot{A}_{n-1})}\right]
\label{SGF}
\eea
As anticipated above, the transformations
induced by the gauge fermions keep producing nondiagonal terms, in which
the physical fields $\s^\a$ mix with the ghosts, the ghosts mix with other
ghosts, and so on, in an infinite sequence. So, at first sight,
in order to obtain the $<\s^\a\bar{\s}^\ad>$ propagator it seems compulsory
to perform a complete diagonalization of the quadratic, kinetic terms in the
quantum action. We want to show now that this infinite series of operations
can actually be avoided and that with few steps one can decouple the $\s^\a$'s
from the ghosts. Since only the
$\s^\a$, $\bar{\s}^\ad$ fields interact with the external background, this
is all we need for the calculation of the one-loop $\b$-function of the
nonlinear $\s$-model. In order to explain the procedure it is simpler to
visualize the kinetic terms
as the infinite sum of the elements of a matrix; moreover to better
distinguish between the various fields it is convenient to use again the
names as they appear in table 1. So we write
\beq
\left( \matrix{ S_{\bar{\s}\s}&S_{\bar \sigma \nu}& 0&0&0&\dots \cr
S_{\bar \nu\sigma }& S_{\bar \nu \nu}& S_{\bar  \nu d}
&S_{\bar \nu \tau}&0&  \dots \cr
0&S_{\bar d \nu}&S_{\bar d d}&S_{\bar{d}\t}&0&\dots \cr
0& S_{\bar \tau \nu}&S_{\bar{\t}d}&S_{\bar \tau \tau}
&S_{\bar \tau f}&\dots \cr
0&0&0&S_{\bar f\tau }&S_{\bar ff}& \dots \cr
\vdots& \vdots&\vdots&\vdots&\vdots&\ddots}
\right)
\label{quadmat}
\eeq
where $S_{\bar x y} \equiv \bar{x} W y$ denotes the quadratic term
between the $\bar{x}$ and $y$ fields.
The disentangling of the $\s^\a$ fields from the rest can be obtained
through the following two successive steps:
first we cancel the $d\n$ terms with a shift
$d\rightarrow d+\n$, thus producing new $\n\n$ and $\n\t$ terms.
Then we simply obtain the final kinetic term for
$\bar{\s} \s$ diagonalizing $\n\s$ with $\n \rightarrow \n+\s$ and
taking advantage of the fact that, contrary to the expectation,
this move does not produce
$\s\t$ terms. At this point the ghosts, while still mixed with each others,
do not have any crossing with the physical $\s^\a$'s.
More specifically, from (\ref{SGF})
we consider the part of the action explicitly indicated in
(\ref{quadmat})
\bea
S_{gf} &=&\bar{\s}^\ad (\Db_\ad D_\a +k^2 D_\a \Db_\ad)\s^\a
+ (k\bar{\s}^\ad D_\a
i \pa^\a_{~\ad}\n +~h.c.)\nonumber\\
&+& \bar{\n} (-2\Box +k^{'2}_1 D_\a
i\pa^{\a\ad} \Db_\ad)\n
+(\bar{d}^{(\ad\bd)}_\a \Db_\bd
i\pa^\a_{~\ad}\n +~h.c.)
+\bar{d}^{(\ad\bd)}_\a (\Db_\bd D_\b +k_1^2 D_\b \Db_\bd)
d^{(\a\b)}_\ad\nonumber\\
&-& (k_1\bar{\t}^\bd_\b
i\pa_\a^{~\ad}\Db_{(\ad} d_{\bd)}^{(\a\b)} +~h.c.)
+(k'_1 k^{''}_1 \bar{\t}^\ad_\a i\pa^\a_{~\ad} \Db^2\n+~h.c.)
+\dots
\label{Sdiag}
\eea
Thus we cancel the $d\n$ crossed terms
with the appropriate shift of the $d$ field
\beq
d_\ad^{(\a\b)}\rightarrow d_\ad^{(\a\b)} +\frac{4(k_1^2-1)}
{3-2k_1^2}  \frac{i\pa^{(\a}_\ad \Db^2 D^{\b)}}{\Box}~ \n
+\frac{2}{3-2k_1^2}\frac{ i\pa^{(\a}_\ad i \pa^{\b)\gd} \Db_\gd }
{\Box}~\n
\eeq
Correspondingly
the additional contributions to the kinetic action of $\n$ are
\beq
\d S_{\bar{\n}\n}= 3 \bar{\n} \left[ D_{\a} \bar{D}^2 D^{\a} +
\frac{2}{3-2k_1^2} D^2 \bar{D}^2 \right] \n
\label{S2n}
\eeq
while the new $\t \n$ crossed terms are
\beq
\d S_{\t\n}=  \frac{3k_1}{3-2k^2_1}\left( \bar{\n}
D^2i\pa_\a^\ad \t^\a_\ad- \bar{\t}^\ad_\a
i\pa^\a_\ad \Db^2\n\right)
\label{xx}
\eeq
At this stage, adding the terms in (\ref{S2n})
to the original quadratic $\bar{\n}\n$ terms in (\ref{nmact2})
and (\ref{nmact3}) we obtain
\beq
 S_{\bar{\n}\n}= \bar{\n}\left[- 2 \Box + \left(\frac{6}{3-2k_1^2}
-2k^{'2}_1\right)D^2 \bar D^2
+(3-k^{'2}_1) D_\a \bar D^2 D^\a \right]\n \equiv \bar{\n}\tilde{W}_\n\n
\label{nuquad}
\eeq
The final step requires the diagonalization of the $\s\n$ terms
(see eq. (\ref{nmact2}))
\beq
\bar{\n}\tilde{W}_\n \n  - k \bar{\n} i \pa_\a^{~\ad} \Db_\ad \s^\a
+k \bar{\s}^\ad D_\a i\pa^{\a}_{~\ad} \n
\eeq
First we easily compute the inverse of the operator in (\ref{nuquad})
\beq
\tilde{W}^{-1}_\n = - \frac{1}{2 \Box} +
\frac{3-3k^{'2}_1+2(k_1 k'_1)^2}
{4k_1^2-6k_1^{'2}+4(k_1 k'_1)^2} \frac{D^2 \bar D^2}{\Box^2}
 + \frac{3-k^{'2}_1}{2(1-k^{'2}_1)} \frac{D_\a \bar D^2 D^\a}{\Box^2}
\label{nuinverse}
\eeq
Then we shift the $\n$ field in such a way to cancel the
$\s\n$ terms
\beq
\n \rightarrow \n+k\tilde{W}^{-1}_\n i\pa_{\a}^{~\ad} \Db_\ad \s^\a
\label{shiftnu}
\eeq
The relevant fact is that (\ref{shiftnu}) does {\em not}
give rise to $\t\s$ contributions: in fact substituting
(\ref{shiftnu}) in (\ref{Sdiag}) and (\ref{xx}) one obtains
\beq
\d S_{\t\s}= k \left( k'_1 k^{''}_1- \frac{3k_1}{3-2k^2_1} \right)
\bar{\t}^\ad_\a
i \pa^\a_{~\ad} \Db^2 \tilde{W}^{-1}_\n
i\pa_{\b}^{~\bd} \Db_\bd \s^\b +~h.c.
\label{tausigma}
\eeq
From the expression of $\tilde{W}^{-1}_\n$ in (\ref{nuinverse})
it is immediate to check that all the terms in (\ref{tausigma})
vanish trivially. Finally no more crossed couplings of the
$\s$-fields are present.

Thus we are left with new
quadratic terms $\bar{\s}^\ad \s^\a$ which combine with the original
$\Db_\ad D_\a +k^2 D_\a \Db_\ad$ and give as total kinetic
operator
\beq
\tilde{W}_{\a \dot \a} =
\bar{D}_{\dot{\a}} D_{\a} + \frac{k^2}{2} D_\a \bar D_{\dot \a} -
 \frac{k^2+(kk^{'}_1)^2}{2-2k^{'2}_1} i \pa_{\a \dot \a}
\frac{D^2 \bar D^2}{\Box}
+ \frac{k^2}{2}i \pa_{\a \dot \a}\frac{D_\b \bar D^2 D^\b}{\Box}
\eeq
The $<\s^\a \bar{\s}^\ad>$ propagator is given by
\bea
<\s^\a \bar{\s}^\ad>&=&
{(\tilde{W}^{-1})}^{\a \dot \a}= - \frac{i \pa^{\a \dot \a}}{\Box}
+\frac{3(kk'_1)^2 + 4 -2k^{'2}_1}{4(kk'_1)^2}
i \pa^{\a \dot \a} \frac{D^2 \bar D^2}{\Box^2}+
\nonumber \\
&&+
\frac{3 k^2 -2}{4 k^2} i \pa^{\a \dot \a} \frac{D_\b \bar D^2 D^\b}{\Box^2}+
\frac{2-k^2}{4 k^2} i \pa^{\a \bd} i \pa^{\b \ad}
\frac{D_\b  \Db_\bd}{\Box^2}
\eea
Now we have collected all the ingredients which are necessary
for the computation of the one-loop $\b$-function for
the linear multiplet $\s$-model.  We follow the same
procedure as in the corresponding $N=2$ chiral multiplet
calculation \cite{b6}.

\section{One-loop beta-function}

Going back to (\ref{actionquad2}) it appears that the
quantum fields $\s^\a$, $\bar{\s}^\ad$ are always coupled
to the external background through their field-strengths
$\S$, $\Sigmab$. This implies that in the perturbative
calculations only the $<\Sigmab\S>$ propagator does enter.
Thus it suffices to consider
\beq
<\Sigmab\S>=D_\a<\s^\a \bar{\s}^\ad>\Db_\ad =
  \frac{ D^2 \bar  D^2}{\Box}+
\frac{D_{ \a}  \Db^2  D^{\a}}{\Box}
\equiv \P
\label{risfin}
\eeq
The above result shows that the {\em effective} propagator is
automatically independent of the gauge parameters
introduced in the gauge-fixing procedure. This provides
a very good check of the methods used in the quantization of
the linear multiplet.

Finally we are ready to compute the one-loop divergence:
the effective Feynman rules can be obtained directly from the
expansion of
\beq
\exp(S_{eff})=
\exp(-\S\P^{-1}\Sigmab+\S {\cal V} \Sigmab +\frac{1}{2} \S {\cal U}\S
+\frac{1}{2} \Sigmab \bar{{\cal U}} \Sigmab )
\eeq
where $\P$ is given in (\ref{risfin}) and (cfr. (\ref{actionquad2}))
\beq
{\cal V}\equiv (F_{\m\bar{\m}}+ \d_{\m\bar{\m}})\qquad,
\qquad{\cal U}\equiv F_{\m\n}
\label{def}
\eeq
The one-loop divergent contributions are computed using standard
$D$-algebra techniques very similar to the ones used in \cite{b6}.
We recall that in order to obtain
local divergent structures, it is sufficient to consider only the
contributions with no derivatives acting on the external background, so that
the covariant spinor
derivatives $D_\a$, $\Db_\ad$ are freely integrated by parts
on the internal quantum
lines of the diagrams and the algebra is easily completed.

We group the various graphs into two sets:
the graphs which contain only $\S {\cal V} \Sigmab$ vertices and
all the others. The first set gives rise to a sum of terms, which
before completion  of the $D$-algebra, we write
schematically in
the form
\beq
-{\rm{tr}}\log(1-\P {\cal V})={\rm{tr}}\sum \frac{1}{n} (\P {\cal V})^n
\eeq
We obtain the relevant, non vanishing
contributions if,
integrating by parts all the covariant
derivatives inside the loop,  we end up with exactly
two $D$'s and two $\Db$'s. Making use of the relation
\beq
\left( \frac{D^2\Db^2+D_\a\Db^2 D^\a}{\Box}\right)^n
= \frac{D^2\Db^2+D_\a\Db^2 D^\a}{\Box}\rightarrow
-\frac{1}{\Box}D^2\Db^2
\eeq
it is fairly straightforward to obtain the one-loop
divergence from this first group of diagrams
\beq
F^{(1)}_1\rightarrow \frac{1}{\epsilon} {\rm{tr}} \sum \frac{1}{n}({\cal V})^n
=-\frac{1}{\epsilon} \rm{tr} \log(1-{\cal V})
\label{F1}
\eeq

Then we consider diagrams which contain a dependence
on the background of the type ${\cal U}$ and $\bar{{\cal U}}$.
These contributions can be efficiently accounted for, defining
first an effective $<\Sigmab\S>$ propagator with the ${\cal V}$-type
vertices resummed. This can be accomplished most easily keeping in mind that,
as mentioned above, we can drop all the terms where covariant spinor derivatives
act on the background fields. Performing explicitly the
sum of all the ${\cal V}$ vertices we obtain
\beq
<<\Sigmab\S>>=\frac{1}{\Box} (D^2\Db^2+D_\a\Db^2 D^\a)\frac{1}
{1-{\cal V}} \equiv \hat{\P}
\eeq
In terms of $\hat{\P}$ the second class of one-loop diagrams can
be written, before
completion of the $D$-algebra, as
\bea
&~&{\rm{tr}} [\frac{1}{2} {\cal U} \hat{\P}\bar{{\cal U}}\bar{ \hat{\P}}
+\frac{1}{4}({\cal U}\hat{\P}\bar{{\cal U}}\bar{ \hat{\P}})^2
+\frac{1}{6}({\cal U}\hat{\P}\bar{{\cal U}}\bar{ \hat{\P}})^3
+\dots]\nonumber\\
&~&~~~~~= \frac{1}{2} {\rm{tr}} \sum_{n=1}^{\infty} \frac{1}{n}
({\cal U} \hat{\P}\bar{{\cal U}}\bar{ \hat{\P}})^n
\eea
In this case the $D$-algebra is performed using
\beq
\left[\frac{(D^2\Db^2+D_\a\Db^2 D^\a)}{\Box}\frac{(\Db^2 D^2+
\Db_\bd D^2 \Db^\bd)}
{\Box}\right]^n
= \frac{D_\a\Db^2 D^\a}{\Box}\rightarrow -2\frac{D^2\Db^2}{\Box}
\eeq
so that the one-loop divergence from the second set of diagrams
is given by
\beq
F^{(1)}_2\rightarrow -\frac{1}{\epsilon} \rm{tr}
\log(1-{\cal U} \frac{1}{1-{\cal V}}
\bar{{\cal U}}\frac{1}{1-{\cal V}})
\label{F2}
\eeq
Adding the results in (\ref{F1}) and (\ref{F2}) we obtain the
total one-loop divergent contribution
\beq
F^{(1)}\rightarrow -\frac{1}{\epsilon} \left[
\rm{tr} \log(1-{\cal V})+\rm{tr}
\log(1-{\cal U} \frac{1}{1-{\cal V}}
\bar{{\cal U}}\frac{1}{1-{\cal V}}) \right]
\label{Ftotal}
\eeq
Using the definitions in (\ref{def}) we rewrite the result
\beq
F^{(1)}\rightarrow -\frac{1}{\epsilon}
\rm{tr} \log[-(F_{\m\bar{\m}}-F_{\m\n}F^{-1}_{\n\bar{\n}}
F_{\bar{\n}\bar{\m}})]
\label{FFtotal}
\eeq

Now we compare the above expression with what expected from duality
correspondence with the nonlinear $\s$-model in terms of chiral
superfields.   The result in (\ref{FFtotal}) has its
counterpart in (\ref{1loopdiv}), where the one-loop divergent
contribution to the \Ka\ potential of the $N=2$ theory is exhibited.
On the other hand in section 1 we have seen that at the level of
the matrices given by the second derivatives of the potentials
the duality transformations  are given in (\ref{dualmetric}). This is
exactly the correspondence established by the results in (\ref{1loopdiv})
and (\ref{FFtotal}).

\section{Conclusions}

We have computed the one-loop $\b$-function of the nonlinear
$\s$-model for nonminimal multiplets and checked that the
result maintains the classical duality properties of the model
with respect to the $N=2$ chiral one.

Our calculation presents a highly nontrivial and concrete
application of the quantization of the complex linear superfield
via the Batalin-Vilkovisky procedure \cite{b3}. It is remarkable that
in this application the relevant sector of the tower of ghosts are
the so-called `extra ghosts' rather than the ghosts or antighosts.
In spite of the presence
of the infinite tower of gauge invariances and of the corresponding
infinite number of ghost fields our result is obtained with no need of formal
manipulations and in a rather cute and straightforward manner.

Several issues deserve now further investigations.

It would be interesting to
extend the analysis to
higher loops and see whether the classical duality transformations remain
unchanged or else, as for bosonic $\s$-models \cite{b7}, perturbative
corrections are induced. Since the $N=2$ chiral $\s$-model
has a vanishing $\b$-function at two and three loops
one could study the corresponding situation for the
model of nonminimal multiplets.

The geometry associated to the chiral $N=2$ theory has been
thoroughly understood and well defined geometrical objects are constructed
in terms of the \Ka\ potential. A complete geometrical interpretation is still
lacking for the dual theory (however see ref. \cite{b8}).

Finally one could consider
mixed models constructed in terms of both chiral and
complex linear superfields, e.g. of the type used for the supersymmetric
description of the low-energy QCD action \cite{b8,b9}, and introduce, in
addition, couplings to the gauge Yang-Mills superfields.

\medskip
\section*{Acknowledgments.}

\noindent
This work was
supported by the European Commission TMR programme
ERBFMRX-CT96-0045, in which S.P., A.R. and D.Z. are associated
to the University of Torino.
\newpage

\appendix
\section{Conventions}

We list here some of the relations involving spinor covariant derivatives
that we have repeatedly used in sections 3 and 4.
\beq
\{D_\a,\Db_\ad\}=i\pa_{\a\ad}
\eeq
\beq
[D_\a,\Db^2]=-i\pa_{\a\ad}\Db^\ad
\eeq
\beq
D^2\Db^2 D^2=\Box D^2
\eeq
\beq
D_\a\Db^2 D^\a=\Db_\ad D^2 \Db^\ad
\eeq
\beq
D_\a i\pa^{\a\ad} \Db_\ad
=-2 D^2\Db^2-D_\a\Db^2 D^\a
\eeq
\beq
\d^{(4)}(\theta-\theta') D^2\Db^2\d^{(4)}(\theta-\theta')=
\frac{1}{2}\d^{(4)}(\theta-\theta')D^\a\Db^2 D_\a \d^{(4)}(\theta-\theta')
=\d^{(4)}(\theta-\theta')
\eeq

\end{document}